\documentclass[lettersize,journal]{IEEEtran}
\usepackage{amsmath,amsfonts}
\usepackage{algorithmic}
\usepackage{array}
\usepackage[caption=false,font=normalsize,labelfont=sf,textfont=sf]{subfig}
\usepackage{textcomp}
\usepackage{stfloats}
\usepackage{url}
\usepackage{verbatim}
\usepackage{graphicx}
\usepackage{pifont}
\usepackage{hyperref}
\usepackage{listings}
\usepackage[normalem]{ulem}
\hypersetup{
    colorlinks=true, 
    linkcolor=black, 
    urlcolor=black,   
    citecolor=black 
}
\def\BibTeX{{\rm B\kern-.05em{\sc i\kern-.025em b}\kern-.08em
    T\kern-.1667em\lower.7ex\hbox{E}\kern-.125emX}}
\usepackage{balance}

\newif\ifblue
\bluefalse 
\newcommand{\added}[1]{\ifblue\textcolor{blue}{#1}\else #1\fi}
\newenvironment{addedblock}{\par\ifblue\color{blue}\fi}{\par}
\newcommand{\deleted}[1]{\ifblue\textcolor{blue}{\sout{#1}}\fi}

\newif\ifblueII
\blueIIfalse 
\newcommand{\addedII}[1]{\ifblueII\textcolor{blue}{#1}\else #1\fi}
\newenvironment{addedblockII}{\par\ifblueII\color{blue}\fi}{\par}

\begin{document}
\title{Text-Queried Audio Source Separation via Hierarchical Modeling}
\author{Xinlei Yin, Xiulian Peng, Xue Jiang, Zhiwei Xiong, Yan Lu
\thanks{Xinlei Yin and Zhiwei Xiong are with the University of Science and Technology of China (email: xyxl\_231829@mail.ustc.edu.cn; zwxiong@ustc.edu.cn).}
\thanks{Xue Jiang is with the School of Information and Communication Engineering, Communication University of China, Beijing 100024, China (e-mail: jiangxhoho@cuc.edu.cn).}
\thanks{Xiulian Peng and Yan Lu are with the Microsoft Research Asia, Beijing 100080, China (e-mail: xipe@microsoft.com; yanlu@microsoft.com).}
\thanks{This work has been submitted to the IEEE for possible publication. Copyright may be transferred without notice, after which this version may no longer be accessible.}
\thanks{This paper is the result of an open-source project starting from Jan. 2024.}}

\markboth{IEEE TRANSACTIONS ON AUDIO, SPEECH, AND LANGUAGE PROCESSING}%
{How to Use the IEEEtran \LaTeX \ Templates}

\maketitle

\begin{abstract}
Target audio source separation with natural language queries presents a promising paradigm for extracting arbitrary audio events through arbitrary text descriptions. Existing methods mainly face two challenges, the difficulty in jointly modeling acoustic-textual alignment and semantic-aware separation within a blindly-learned single-stage architecture, and the reliance on large-scale accurately-labeled training data to compensate for inefficient cross-modal learning and separation. To address these challenges, we propose a hierarchical decomposition framework, \textit{HSM-TSS}, that decouples the task into global-local semantic-guided feature separation and structure-preserving acoustic reconstruction. Our approach introduces a dual-stage mechanism for semantic separation, operating on distinct global and local semantic feature spaces. We first perform global-semantic separation through a global semantic feature space aligned with text queries. A Q-Audio architecture is employed to align audio and text modalities, serving as pretrained global-semantic encoders. Conditioned on the predicted global feature, we then perform the second-stage local-semantic separation on AudioMAE features that preserve time-frequency structures, followed by semantic-to-acoustic reconstruction. We also \deleted{propose an instruction processing pipeline to parse arbitrary }\added{split} text queries into structured operations, \textit{extraction} or \textit{removal}, coupled with audio descriptions, enabling \deleted{flexible}\added{bidirectional} sound manipulation. Our method achieves state-of-the-art separation performance with data-efficient training while maintaining superior semantic consistency with queries in complex auditory scenes.

\end{abstract}


\section{Introduction}

\begin{figure}[t]
    \centering
    \includegraphics[width=\columnwidth]{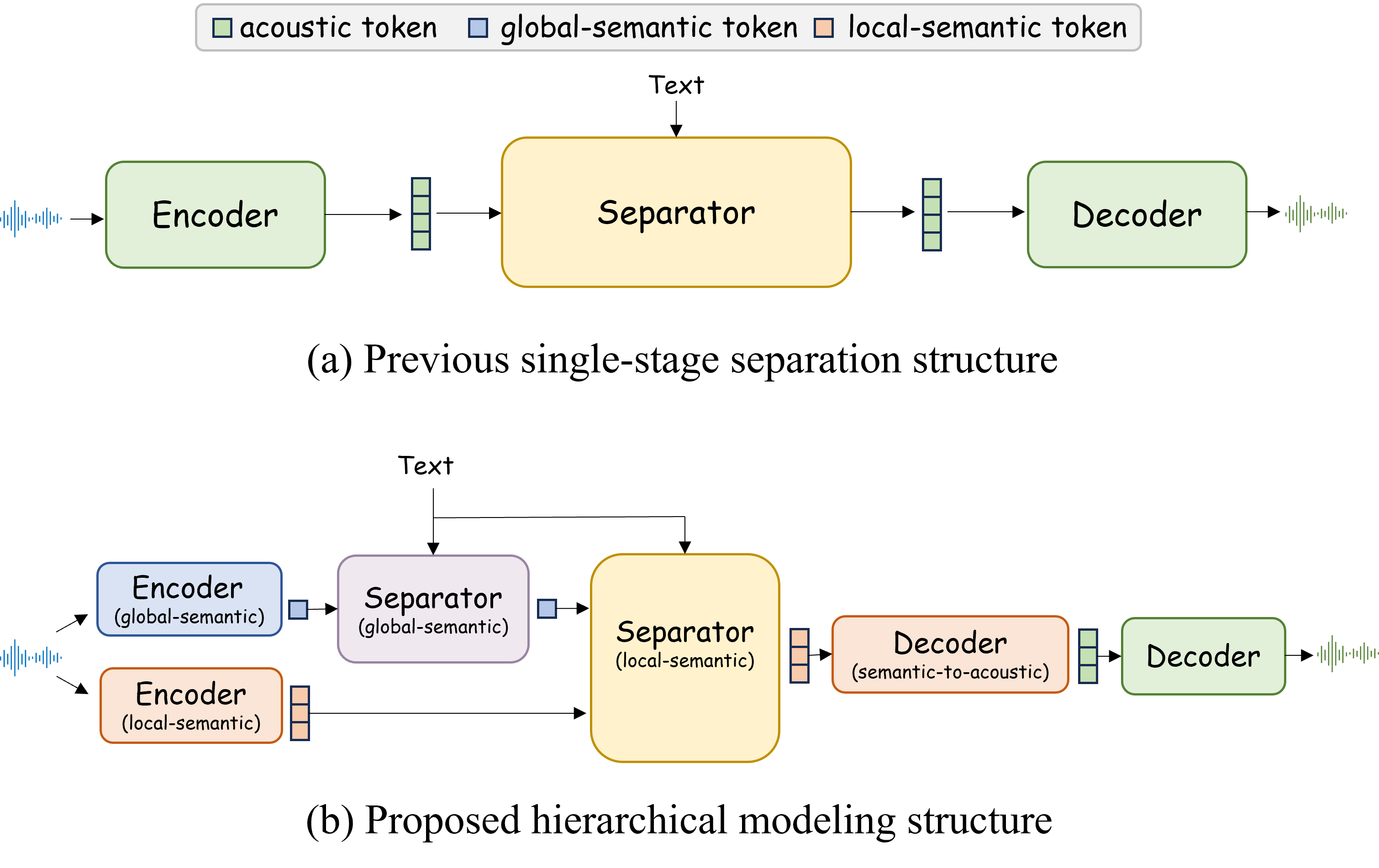}
    \caption{Comparison between previous blind single-stage and our hierarchical modeling frameworks.}
    \hspace{-0.4cm}
    \label{fig:compare}    
\end{figure}

\IEEEPARstart{R}{eal}-world environmental sounds typically comprise diverse audio events from multiple sources. Target sound separation, which isolates specific sound components from mixtures across domains like speech \cite{divide_conquer,casa,luo2019conv}, general audio \cite{liu2024separate}, and music \cite{Kong2021DecouplingMA}, conventionally relies on single-source training samples and focuses on separating predefined source types \cite{wang2018supervised}. Recent advances in universal sound separation (USS) \cite{kong2023universal} have expanded this capability to arbitrary sound sources in real-world recordings. However, the inherent complexity of separating overlapping audio events with varying characteristics persists as a fundamental challenge. Text-queried target sound extraction (TSE), which uses natural language descriptions to selectively separate sounds, has emerged as a promising solution. Unlike audio-visual \cite{zhao2018sound,tzinis2020into} and audio-queried \cite{kong2023universal,chen2022zero,gfeller2021one} methods, it provides greater flexibility in query formulation. It also outperforms label-queried \cite{veluri2023real,delcroix2022soundbeam,wang2022improving} methods by eliminating the need for fixed label categories, thereby supporting queries of any type and facilitating open-domain generalization.

The principal challenge in text-queried target sound extraction lies in establishing robust cross-modal associations between linguistically diverse queries and intricate acoustic patterns. Natural language instructions may contain temporally ordered event sequences (e.g. ``A man talks to himself happily before playing the violin.") or detailed sound characteristics (e.g. heavy rain, high/low frequency engine), requiring precise alignment with corresponding audio segments in potentially overlapping mixtures.  

Recently, an increasing number of studies have explored this task, which can mainly be divided into two categories: mask-based separators \cite{liu2022separate,dong2022clipsep,mahmud2024weakly,liu2024separate, cheng2024omnisep} and conditional generative models \cite{wang2023audit,yang2023uniaudio,han2024instructme,yuan2025flowsep}. The dominant mask-based approaches typically employ separation networks to estimate a multiplicative time-frequency mask through conditioned U-Net architectures. These methods incorporate a conditioning signal into the audio U-Net with a query network to predict the target mask. The other stream, the emerging generative paradigm by transformer \cite{kreuk2022audiogen}, diffusion \cite{wang2023audit} or \added{flow matching} \cite{yuan2025flowsep} models, formulates separation as a specialized audio editing task \cite{wang2023audit}. 

Despite their varied frameworks, these methods all employ a joint optimization of cross-modal understanding and acoustic separation within a single-stage end-to-end framework (shown in Figure \ref{fig:compare} (a)). This predominantly data-driven approach could potentially lead to \deleted{training instability and }an increased risk of overfitting, necessitating the use of large-scale audio-text datasets. The fundamental limitations persist when operating on spectrograms \added{capturing acoustic information} \cite{liu2024separate} or VAEs' latent \cite{wang2023audit} feature, \added{as they lack explicit semantic representation \cite{liu2024audioldm}}. This forces models to resolve ambiguities through acoustic details alone, often resulting in under-separation or inconsistent audio events. Furthermore, commonly used weakly-labeled datasets like AudioSet \cite{gemmeke2017audio} and VGGSound \cite{chen2020vggsound} contain significant label noise and irrelevant audio content, which can confuse the trained models and further complicate text instruction understanding. 

In this work, we decompose the text-queried audio separation into semantic-feature-level manipulation at different levels followed by semantic-to-acoustic decoding and acoustic reconstruction. Different from acoustic representations (e.g., spectrograms or VAE latents), we attempt to operate separation in a more compact global-local semantic feature space. The global semantic space aligns audio and text, inside which we can extract target semantic hints. In the local semantic space that preserves not only event semantics but also local time-frequency details for reconstruction, we can extract more fine-grained semantic features conditioned on previous-stage predicted global hints. Such a fine-grained feature is then utilized to reconstruct the waveform through a semantic-to-acoustic generative model. This hierarchical modeling with dual-semantic-stage separation and semantic-to-acoustic reconstruction (shown in Figure \ref{fig:compare} (b)) effectively separates target audio events that align well with the text query without using large weakly-labeled datasets. \added{In addition, we formulate target audio source separation as two complementary tasks, enabling the extraction of target sources and the removal of interferences. For instance, the query ``\textit{Remove the dog barking}'' can achieve the same goal as ``\textit{Extract the rain sound}'' in a mixture if ``\textit{rain}'' is the desired sound. }

\added{In this paper, \textit{semantic} refers to high-level interpretations of audio and text, such as sound event categories or descriptive labels. \textit{global semantic} features capture holistic, event-level information like the overall context or type of sound, without acoustic details such as volume or timbre. In contrast, \textit{local semantic} features represent more fine-grained, time-frequency-aware detailed semantic information, such as specific words in speech or the timing and location of sound events, capable to reconstruct the waveform.}

Our contributions can be summarized as follows:
\begin{enumerate}
  \item{We propose a hierarchical modeling paradigm, \textit{HSM-TSS}, to separate audios on different feature levels queried by text. By decomposing global and local semantic representations with decoupled semantic separation and acoustic reconstruction, the proposed method alleviates the challenge of learning audio-text alignment and extracting semantically-correct audio events simultaneously, effectively \deleted{increasing the data efficiency and}\added{enhancing} our model generalizability.}  
  \item{We pretrain a text-audio aligned audio representation, \textit{Q-Audio}, through contrastive learning, which outperforms the commonly used CLAP \cite{wu2023large, elizalde2023clap} in several benchmarks\added{, including AudioCaps and Clotho for both text-to-audio retrieval and audio captioning.}}
  \item{We conduct experiments on several datasets \added{including Clotho, AudioCaps, and FSD50K} and demonstrate that the proposed hierarchical modeling achieves state-of-the-art separation performance by both signal-level and semantic-level metrics.}
\end{enumerate}

\section{Related work}
\subsection{Unconditional sound separation}
Unconditional audio separation aims to decompose mixed audio signals into individual sources without relying on external queries or prior knowledge. Early research primarily focuses on domain-specific tasks, such as speech separation \cite{wang2018supervised, luo2019conv} and music source separation \cite{takahashi2021densely}. These methods address label ambiguity problem by permutation invariant training (PIT) \cite{yu2017permutation} that permutes predictions to match target signals during training, and often require post-processing to select target sources from separated outputs. However, they typically rely on single-source training data, limiting the scalability to real-world mixtures. To mitigate this, mixture invariant training (MixIT) \cite{wisdom2020unsupervised} introduces unsupervised learning with multi-source mixtures. However, it tends to over-separate and requires post-selection. Subsequent work combines MixIT with pre-trained sound classifiers \cite{wisdom2021sparse}, yet these classifiers still require single-source annotations. Other approaches, like MixPIT \cite{karamatli2022mixcycle}, attempted direct prediction from mixtures but encountered under-separation problems. Weakly supervised methods \cite{pishdadian2020finding} utilized sound classifiers but were constrained by fixed label categories. These limitations underscore the challenge of achieving open-domain separation without single-source supervision or post-processing.

\subsection{Text-queried sound separation}
Text-queried audio separation leverages natural language descriptions or labels to guide the extraction of target sounds from mixtures. Early methods explored label-queried separation \cite{veluri2023real, delcroix2022soundbeam}, which relies on predefined class labels and struggles with generalization to unseen categories. A paradigm shift emerged with language-queried audio source separation (LASS), which allowed for flexible separation using free-form text queries. Pioneering efforts like LASS-Net \cite{liu2022separate} introduced end-to-end training with audio-text pairs, while CLIPSep \cite{dong2022clipsep} utilized contrastive language-image pretraining (CLIP) \cite{radford2021learning} to align text and audio embeddings through a visual modality, achieving zero-shot separation. \added{AudioSep \cite{liu2024separate} scaled training with large audio-language datasets and showcased robust open-domain generalization. OmniSep \cite{cheng2024omnisep} further enabled separating target audios through both single-modal and multi-modal composed queries. Some recent advancements have leveraged audio‑only data, i.e., without paired captions, to mitigate the lack of annotations for text‑queried extraction \cite{saijo2025leveraging, ma2025language}. Researchers have also been exploring audio separation with generative models like diffusion models \cite{hai2024dpm} and rectified flow matching \cite{yuan2025flowsep}.}\deleted{Recent advancements have incorporated multi-modal supervision \cite{tan2023language, cheng2024omnisep} and hybrid training \cite{kilgour2022text},to tackle data scarcity. For instance, AudioSep \cite{liu2024separate} scaled training with large audio-language datasets and showcased robust open-domain generalization.} Despite these advancements, challenges remain in handling linguistic diversity (e.g., paraphrased descriptions) and noisy real-world data. Alternative approaches, like audio-visual separation \cite{gao2019co, tzinis2020into}, used visual cues but were sensitive to occlusions or off-screen sounds. Conversely, text-based methods possess broader applicability, as natural language offers a scalable and intuitive interface for specifying target sources.

\subsection{Audio editing with generative models}
The most common approach for audio editing is to train specialized models for particular tasks, like style transfer \cite{wu2023musemorphose,popov2021diffusion} or inpainting \cite{borsos2022speechpainter,marafioti2020gacela}. Recently, some works have studied general audio editing following human instructions. AUDIT \cite{wang2023audit} designs several editing tasks involving adding, removal, inpainting, replacement, and super-resolution based on latent diffusion models. UniAudio \cite{yang2023uniaudio} is an audio foundation model which takes adding,
dropping and super-resolution as general audio editing tasks. In terms of this, target sound separation can be seen as a sub-task of instruction-based audio editing. However, they still leveraged common acoustic representations, like VAEs' latent or neural codec tokens, and typically suffered from under-separation or over-separation problems. Instead, our approach investigates the hierarchical semantic modeling paradigm with pretrained representations at different feature levels, with which it achieves superior performance with carefully synthesized training data with accurate labels.      

\vspace{-2 mm}
\subsection{Audio representation learning}
Recent advancements in self-supervised learning (SSL) have significantly shaped audio representation learning, empowering models to extract meaningful features from raw audio data without labeled supervision. These approaches can be broadly categorized into predictive, contrastive, and masked predictive modeling techniques, each addressing distinct challenges in audio processing. Predictive modeling approaches, such as autoregressive predictive coding (APC) \cite{chung2020generative,chung2020vector}, have been pivotal in learning sequential audio representations by predicting future elements of a sequential audio input based on past data. 
In masked predictive approach, MAE-AST \cite{baade2022mae} adapted the masked autoencoders (MAEs) \cite{he2022masked} for audios, combining discriminative and generative objectives for training. It splits audio spectrogram into patches, mask a portion of them, and use transformers to reconstruct the masked segments. The work AudioMAE \cite{huang2022masked}, further investigated shifting windows and local self-attention mechanisms to enhance modeling. 
Unlike predictive and masked approaches, contrastive learning has largely been utilized to learn cross-modal representations such as CLIP \cite{radford2021learning} in computer vision. It has also been adapted for audio , as seen in CLAP \cite{elizalde2023clap,wu2023large}, which unifies general audio and text into a joint embedding space. These multi-modal approaches provide weak supervision for tasks like multi-source sound separation, further enriching audio representation learning. Motivated by BLIP-2 \cite{li2023blip} that leverages a Q-Former for vision-language representation learning, we introduce a lightweight module Q-Audio as a bridge between text and audio with frozen audio and text encoders in our hierarchical modeling. \added{Similar Q-Former-style bridging between speech/audio and text has also been explored in prior work, e.g. ZeroST \cite{khurana2024zerost} for zero‑shot speech translation.}

\section{The proposed framework}


\begin{figure*}[t]
    \centering
    \includegraphics[width=\textwidth]{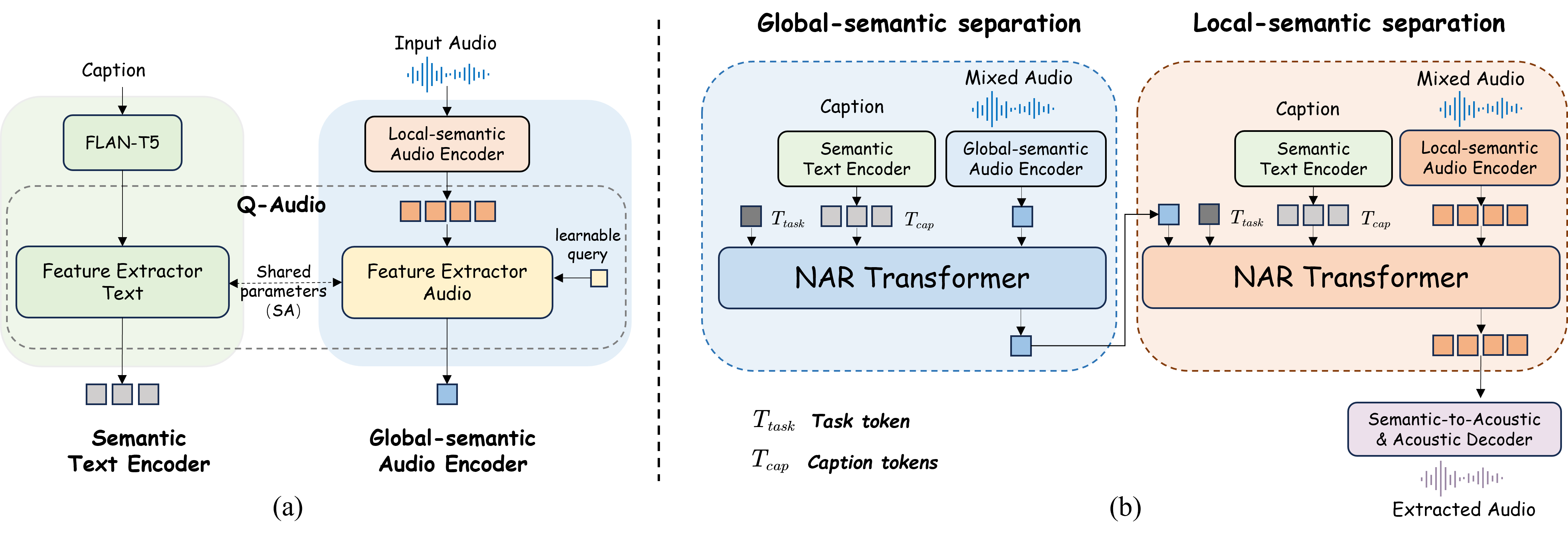}
    \caption{Overview of our proposed hierarchical modeling and separation frameworks. (a) Two-level audio feature representation.  (b) Text-queried two-level separation and decoupled semantic separation and acoustic reconstruction.}
    \label{fig:inference_pipeline}
\end{figure*}

\subsection{Overview}
Let $x_{mix} \in \mathbb{R}^{L}$ denote the audio mixture of the target audio $x_{tgt} \in \mathbb{R}^{L}$ and an interfering audio $x_{other} \in \mathbb{R}^{L}$, given by
\begin{equation}
x_{mix} = x_{tgt} + x_{other}.
\end{equation}
$L$ is the audio length. The goal is to separate the target audio source $x_{tgt}$ based on a given text instruction $\mathcal{T}$. This separation process can be formalized as $\mathcal{H}: (x_{mix}, \mathcal{T}) \mapsto x_{tgt}$, where $\mathcal{H}$ denotes the separation model. Conventional TSE methods \cite{liu2022separate,dong2022clipsep,mahmud2024weakly,liu2024separate} typically rely on a purely data-driven approach through a single-stage end-to-end blind learning architecture, where semantic and acoustic information are entangled in the mixture and it struggles to distinguish target audio events from the interference. Inspired by audio and speech generative models \cite{liu2024audioldm, borsos2023audiolm}, we reformulate the TSE task as a regeneration process by $\mathcal{H}: (x_{mix}, \mathcal{T}) \mapsto S \mapsto x_{tgt}$. Here, an intermediate feature $S$ is introduced to decouple semantic separation with acoustic reconstruction. Such an intermediate representation $S$ is both semantic and information-rich in containing local semantics to reconstruct the target audio. We take it as \textit{local} semantic feature in our approach. 

To provide more robust semantic guidance, we further decompose the semantic feature $S$ by extracting a semantic-only \textit{global} feature $G$ on top of it, which captures high-level audio event descriptions (e.g., \textit{dog barking}) without spatial or acoustic details.\deleted{and aligns well with the text feature space. }\added{Here, spatial/acoustic details refer to fine-grained characteristics of the audio signal (e.g. spatial cues or timbral nuances) that are typically not fully described in the accompanying text. This design ensures that $G$ aligns well with the text feature space.} Comparatively, $S$ is more fine-grained, which integrates both semantic representation (e.g. \textit{what happened in the audio}) and acoustic properties (e.g., \textit{pitch or spatial positioning}) necessary for waveform reconstruction. This motivates a hierarchical framework that \added{not only decouples semantic separation with acoustic reconstruction but also} divides the feature-level separation into \textit{global-semantic} and \textit{local-semantic} stages, with the predicted global feature $\hat{G}$ acting as a conditioning input to refine the extraction of the local feature $\hat{S}$. The decomposition process can be denoted as $\mathcal{H}: (x_{mix}, \mathcal{T}) \mapsto G\mapsto S \mapsto x_{tgt}$. 

This hierarchical design generates audio representations across three distinct levels: the global-semantic level (via $\hat{G}$), the local-semantic level (via $\hat{S}$), and the acoustic-only level (via $x_{tgt}$). Consequently, our proposed framework is structured into three sequential components: 
\begin{equation}
    \mathcal{H}_1: (x_{mix},\mathcal{T}) \mapsto \hat{G}, 
\end{equation}
\begin{equation}
    \mathcal{H}_2: (x_{mix},\mathcal{T}, \hat{G}) \mapsto \hat{S},
\end{equation}
\begin{equation}
    \mathcal{H}_3: \hat{S} \mapsto x_{tgt}.
\end{equation}
$\mathcal{H}_1$, $\mathcal{H}_2$, $\mathcal{H}_3$ denote global-semantic separation, local-semantic separation and \added{semantic-to-acoustic and} acoustic decoders, respectively. As depicted in Figure \ref{fig:inference_pipeline} (b), each stage yields a specialized representation tailored to its role in the pipeline. The semantic-to-acoustic and acoustic decoder maps the local-semantic feature $\hat{S}$ to the acoustic domain and then reconstructs the target waveform using a neural codec decoder. In the following sections \ref{sec:stage1}, \ref{sec:stage2} and \ref{sec:decoder}, we will describe these three modules in detail.

\deleted{To enhance the model’s ability to interpret diverse instructions, we leverage a pre-trained LLM to }\added{We }decompose the text input into two components: the task type $T_{task}$ and the caption $T_{cap}$, as shown in the input of Figure \ref{fig:inference_pipeline}. By isolating the captions from task types, we can exploit the strengths of audio-language contrastive learning that aligns text and audio in a shared representation space. We define two separation tasks: \textit{removal} and \textit{extraction}, allowing the captions to describe either the audio events to be removed or those to be retained\deleted{, thereby increasing the flexibility of user instructions}. This process is elaborated in Section \ref{sec:instruction}.

Compared to single-stage methods, this three-level hierarchical framework offers several advantages. It disentangles semantic and acoustic processing to mitigate the impact of interference, enables modular optimization of each stage, and improves robustness in challenging scenarios where target and interfering audio events overlap significantly \added{as evidenced in the experimental part}. By integrating instruction decomposition with staged feature extraction, our approach achieves precise and flexible text-queried sound separation.

\subsection{Two-level semantic representations}

\subsubsection{Local semantic representation}\label{sec:audiomae}
We leverage the AudioMAE \cite{huang2022masked}, a self-supervised learned audio representation, for the local semantic feature $S$. Unlike discrete neural codec codes that prioritizes acoustic fidelity but lacks semantic representativeness \cite{borsos2023audiolm, jiang2024universal}, or contrastively learned cross-modal features CLAP that lose details, AudioMAE balances acoustic details and semantic information by predicting masked acoustic features. This makes it a good local-semantic representation, clustering semantically similar audio events while retaining spatial-temporal structure, \added{as also discussed in AudioLDM 2 \cite{liu2024audioldm}}. 

AudioMAE takes a Vision Transformer (ViT)-based encoder-decoder structure. It takes the log mel spectrogram \(X \in \mathbb{R}^{T \times F}\) of an audio signal \(x\) as input, splits it into \(P \times P\) patches, and embeds them into a \(C_s\)-dimensional latent space, yielding features of a shape \(\frac{T}{P} \times \frac{F}{P} \times C_s\). \(T\) and \(F\) are the number of frames and mel-frequency bins, respectively. Its masked autoencoder design uses an asymmetric structure, pairing large encoders with small decoders and scales well for linear probing, thanks to a high masking ratio that reduces encoding complexity \cite{he2022masked}. Following MW-MAE \cite{yadav2023masked}, we adopt a small 4-layer transformer-based decoder with a width of 384 and 8 attention heads and pretrain it on AudioSet \cite{gemmeke2017audio}. 

\subsubsection{Global semantic representation}\label{sec:qformer}
While AudioMAE captures both semantic and acoustic details, it lacks alignment with text feature space, limiting cross-modal understanding. To address this, we introduce \textit{Q-Audio} as the cross-modal global semantic feature extractor, inspired by BLIP-2 \cite{li2023blip}. As depicted in Figure \ref{fig:inference_pipeline} (a), Q-Audio bridges features from a frozen local semantic audio encoder (our pretrained AudioMAE) and a frozen FLAN-T5 text encoder \cite{chung2024scaling} into a shared space. It comprises two feature extractors that utilize a transformer structure and share self-attention layers. The audio part extracts semantic components from AudioMAE via a self-attention followed by a cross-attention, and the text feature extractor utilizes only self-attention layers. \deleted{We adopt one learnable query embedding to interact with the frozen audio features through cross-attention layers.}\added{We adopt one learnable query embedding, a trainable latent vector that selectively aggregates and distills information from input representations \cite{li2023blip}. This embedding interacts with the frozen audio features via cross-attention mechanisms, thereby facilitating efficient extraction of global semantic cues aligned with the textual query.} The query also interacts with the text branch in self-attention layers through masking. This structure extracts global semantic information from the local-semantic representation while alleviating text comprehension demands through FLAN-T5, which is different from the original BLIP-2 design \cite{li2023blip}. This design makes the extractor an audio-language aligner with a much lower complexity than the original one. What's more, extracting global feature $G \in \mathbb{R}^{1 \times C_{g}}$ from the local-semantic representation $S \in \mathbb{R}^{\frac{T}{P} \times \frac{F}{P} \times C_s}$ further ensures the correlation and consistency between two-level representations, better facilitating our hierarchical modeling and separation design. 

During the learning phase, the main objective is to enable the learnable query to extract audio representations that are most informative for the corresponding text input. We adopt the original optimization design in \cite{li2023blip} with three training objectives: audio-language contrastive learning, audio-language matching, and audio-grounded text generation, each leveraging a different attention masking mechanism to regulate the interaction between the learnable query and the text branch in self-attention layers. As shown in Figure \ref{fig:inference_pipeline}, we use the audio branch of Q-Audio as the global-semantic audio encoder, and the text branch as the text encoder for the two-level separation. 

\subsection{Global-semantic separation}\label{sec:stage1}
The first stage $\mathcal{H}_1: (x_{mix},\mathcal{T}) \mapsto \hat{G}$ of our separation framework operates in the Q-Audio feature space to separate semantic representations of the target audio. We use Q-Audio as the mixture audio and text encoders and employ a 6-layer non-autoregressive (NAR) transformer to predict the target semantic feature $G_{tgt} \in \mathbb{R}^{1 \times C_{g}}$. \added{Here, $G_{tgt}$ denotes the global semantic feature extracted by Q-Audio taking the ground-truth target signal $x_{tgt}$ as input.} This prediction is conditioned on the concatenation of the mixture audio feature $G_{mix} \in \mathbb{R}^{1 \times C_{g}}$, the text feature $T_{cap} \in \mathbb{R}^{N \times C_{t}}$, and the task token $T_{task}$. $N$ is the sequence length of text tokens and $C_{t}$ is the token dimension. The semantic alignment between audio and text feature inputs enables the transformer to accurately model audio-text relationships and guide the separation process. 

The optimization involves two objectives: a similarity loss and an L1 loss. The similarity loss is defined as the cosine similarity between target $G_{tgt}$ and predicted features $\hat{G}$, which is given by
\begin{equation}
    \mathcal{L}_{sim} = 1 - \cos(\hat{G}, G_{tgt}).
\end{equation}
The L1 loss is the L1-norm distance between two features given by
\begin{equation}
    \mathcal{L}_{L1} = ||\hat{G} - G_{tgt}||_{1}.
\end{equation}
The total loss is a weighted combination of these two terms with weights $\lambda_1$ and $\lambda_2$, ensuring both vector-wise and element-wise accuracy. It is defined as follows
\begin{equation}
    \label{eq:stage1_loss}
    \mathcal{L}_{global} = \lambda_1\mathcal{L}_{sim}+\lambda_2\mathcal{L}_{L1}.
\end{equation}
The predicted $\hat{G}$ serves as a semantic guidance for the subsequent local-semantic separation stage. 

\subsection{Local-semantic separation}\label{sec:stage2}
The second stage $\mathcal{H}_2: (x_{mix},\mathcal{T}, \hat{G}) \mapsto \hat{S}$ operates in the more fine-grained AudioMAE feature space. This stage takes the mixture audio feature $S_{mix} \in \mathbb{R}^{\frac{T}{P} \times \frac{F}{P} \times C_s}$ by AudioMAE encoder as input to predict the target \added{AudioMAE} feature $S_{tgt} \in \mathbb{R}^{\frac{T}{P} \times \frac{F}{P} \times C_s}$ \added{of the target audio $x_{tgt}$}. The previous-stage output $\hat{G}$, text feature $T_{cap} \in \mathbb{R}^{N \times C_{t}}$, and the task token $T_{task}$ are concatenated as the conditioning input, as shown in Figure \ref{fig:inference_pipeline}. During training, ground-truth global audio feature $G_{tgt}$ is used instead of the prediction $\hat{G}$. 

We leverage the L1-based regression loss for this stage optimization, which is
\begin{equation}
    \label{eq:stage2_loss}
    \mathcal{L}_{local} = ||\hat{S}-S_{tgt}||_{1},
\end{equation}
Where $\hat{S}$ is the predicted output. 
To mitigate possible error propagation between two stages, we further conduct joint fine-tuning by Equations \ref{eq:stage1_loss} and \ref{eq:stage2_loss} after optimizing each stage independently. We introduce a switcher to control the conditioning input of the second stage, selecting ground-truth $G_{tgt}$ with a probability of $P_{gt}$, and prediction $\hat{G}$ with a probability of $P_{pred} = 1-P_{gt}$, \added{similar to that in AudioLDM 2 \cite{liu2024audioldm}}. This joint fine-tuning optimizes the two stages end-to-end, minimizing inconsistencies and improving separation accuracy.

\subsection{Semantic-to-Acoustic and Acoustic decoder}\label{sec:decoder}
After the two-stage separation, the last part $\mathcal{H}_3: \hat{S} \mapsto x_{tgt}$, the decoder, aims to reconstruct the waveform from the predicted local semantic feature. In light of the power of generative models, we leverage auto-regressive transformer for this stage. Specifically, we encode $x_{tgt}$ into discrete acoustic tokens $A$ using a neural audio codec, and leverage an autoregressive transformer to generate audio tokens conditioned on $\hat{S}$, followed by a neural codec decoder to reconstruct the waveform, similar to that in UniCodec \cite{jiang2024universal}.

\subsubsection{Autoregressive audio token generation}
\begin{figure}[t]
    \centering
    \includegraphics[width=0.95\columnwidth]{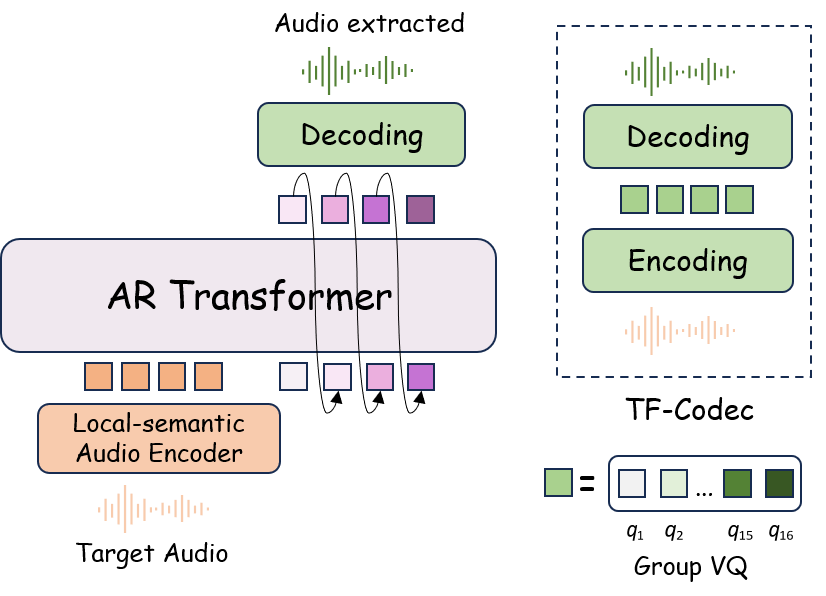}
    \caption{Overview of the semantic-to-acoustic and acoustic decoder. The autoregressive transformer generates acoustic tokens by our neural codec TF-Codec, conditioned on local semantic features.}
    \label{fig:stage3}
\end{figure}

As illustrated in Figure \ref{fig:stage3}, an autoregressive transformer decoder is employed to convert local semantic tokens into acoustic tokens, given by
\begin{equation}
    p(A|S_{tgt};\theta)=\prod_{t=0}^{T_a}p(a_{t}|a_{<t},S_{tgt};\theta),
    \label{eq:ar}
\end{equation}
where $T_a$ denotes the acoustic token sequence length, and $a_{t}$ is the $t$-th frame token of $A$. $a_{-1}$ and $a_{T_a}$ are start and end tokens, respectively. $\theta$ denotes the network parameters. The decoder is trained using a teacher-forcing approach with a cross-entropy loss by
\begin{equation}  
    \begin{aligned}
        \mathcal{L}_{ce} &= -\log \prod_{t=0}^{T_a} p(a_{t}|a_{<t}, S_{tgt};\theta).
    \end{aligned}   
\end{equation}
The generated tokens are then used by the neural codec decoder to reconstruct waveforms. During inference, predictions are made token by token, using each predicted token as input for the next. 

\subsubsection{Reconstruction via TF-Codec}

TF-Codec \cite{jiang2023latent} is a low-latency neural speech codec designed for high-quality speech at low bitrates. We retrain a non-predictive version of TF-Codec to adapt to general audio for acoustic tokens. It employs a VQ-VAE framework, including an encoder, a group vector quantizer, and a decoder optimized end to end. To ensure high perceptual quality for diverse general audio, we apply adversarial training with a multi-scale mel discriminator \cite{yang2023uniaudio} to replace the original single-scale frequency-domain discriminator in \cite{jiang2023latent}. Other training losses are the same as that in \cite{jiang2023latent}. Instead of training from scratch, we finetune from the pre-trained TF-Codec speech codec model for better performance. Similar to UniCodec \cite{jiang2024universal}, in autoregressive token generation of Equation \ref{eq:ar}, all groups of $t$-th step are simultaneously generated in a single stage, leading to a short token length. 

\subsection{Bidirectional Query Instruction}\label{sec:instruction}
\deleted{In text-guided audio separation, users typically provide descriptive prompts to retain some audio components or remove some undesired audio events. This instruction is arbitrary and may target at single or multiple sound sources with specified characteristics. To facilitate this open-vocabulary text prompts with bidirectional operations, our framework introduces dual separation operations: removal and extraction, allowing the model to isolate target audio through complementary approaches. We also utilize pre-trained LLMs to parse text prompts into this task type and a target audio description. For example, an input ``Could you help me separate the sound of a dog barking and the background music?" is decomposed into a task type ``extraction" and an audio description ``a dog barking and the background music", enabling distinct processing of operation intent and acoustic context.}

\deleted{This process offers two critical advantages. It enhances the prompt understanding capability of the separation model. What's more, the decomposition of event description from operation types makes it feasible to leverage audio-language contrastive learning to extract aligned audio-text features, improving the separation model’s ability to capture cross-modal correlations. We introduce a dual-channel text conditioning mechanism for this decomposed processing. The task type is encoded using pretrained token representations from T5’s vocabulary, producing continuous task tokens $T_{task} \in \mathbb{R}^{L\times C_t}$, where $L$ is the token length and $C_t$ is the embedding dimension. Simultaneously, the event description is processed by the semantic text encoder, denoted as $T_{cap} \in \mathbb{R}^{N \times C_t}$, where $N$ is the length of the caption sequence. These two streams of tokens are concatenated along the sequence dimension to form the final text conditioning input $[T_{task},T_{cap}] \in \mathbb{R}^{(L+N) \times C_t}$, as shown in Figure \ref{fig:inference_pipeline} (b). }

\begin{figure}[t]
    \centering
    \includegraphics[width=0.9\columnwidth]{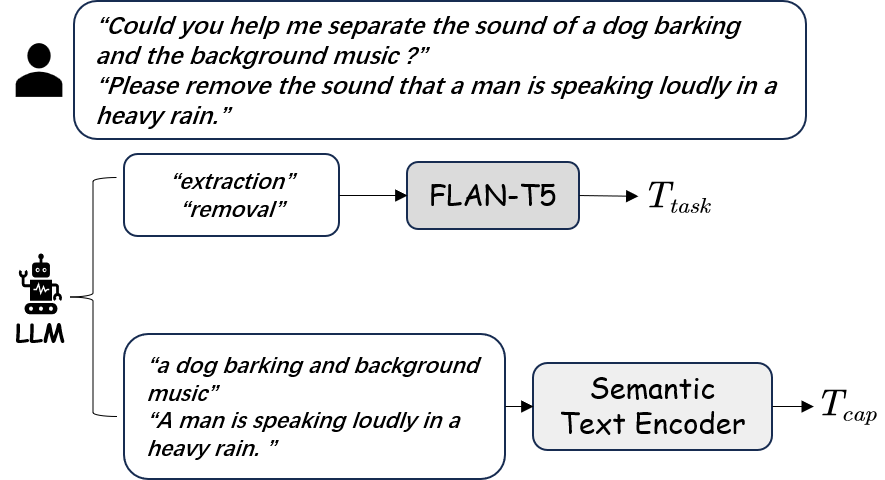}
    \caption{The pipeline of processing arbitrary text instructions.}
    \label{fig:instruction}
\end{figure}

\begin{addedblock}
In text-guided audio separation, users typically provide descriptive prompts to retain certain audio components or remove undesired audio events. To support bidirectional operations, our framework formulates the separation task into two complementary modes: \textit{removal} and \textit{extraction}. Each instruction is decomposed into a task type, indicating the operation, and an audio event description, specifying the relevant sound sources.

We introduce a dual-channel text conditioning mechanism for this decomposed processing. The task type is encoded using pretrained token representations from T5’s vocabulary, producing continuous task tokens $T_{task} \in \mathbb{R}^{L\times C_t}$, where $L$ is the token length and $C_t$ is the embedding dimension. Simultaneously, the event description is processed by the semantic text encoder, denoted as $T_{cap} \in \mathbb{R}^{N \times C_t}$, where $N$ is the length of the caption sequence. These two streams of tokens are concatenated along the sequence dimension to form the final text conditioning input $[T_{task},T_{cap}] \in \mathbb{R}^{(L+N) \times C_t}$, as shown in Figure \ref{fig:inference_pipeline} (b).

This design offers two critical advantages. First, it enhances the query understanding capability and extensibility of the separation model by structurally disentangling the operation type from the audio event description. Second, this decomposition enables the use of audio-language contrastive learning to extract aligned audio-text features, improving the model’s ability to capture cross-modal correlations.
In practical cases, for arbitrary instructions (e.g. \textit{Can you help me separate the sound of a dog barking and the background music?}), we can utilize large language models (LLM) to assist in parsing natural language prompts into the structured format by our framework, as shown in Figure \ref{fig:instruction}.

\end{addedblock}


\section{Experimental setting}
\subsection{Dataset and preprocessing}

\subsubsection{Semantic separation}\label{sec:semantic extraction}
We utilize audio data from AudioCaps \cite{kim2019audiocaps}, ESC-50 \cite{piczak2015esc}, Clotho v2 \cite{drossos2020clotho}, FSD50K \cite{fonseca2021fsd50k} and WavCaps \cite{mei2024wavcaps} due to their relatively accurate and diverse labels or captions. Each audio contains single or multiple audio events with or without overlaps. AudioCaps is a subset of AudioSet \cite{gemmeke2017audio} with handcrafted captions and we use 45K audio clips. Clotho v2 provides human-annotated captions, and we use 3,839 training clips, each accompanied by five captions. ESC-50 comprises 2,000 environmental audio recordings, evenly distributed across 50 semantic classes, including natural sounds, non-speech human sounds, domestic sounds, and urban noises. FSD50K contains approximately 40K audio clips with 200 classes, varying in duration from 0.3 to 30 seconds. WavCaps includes four subsets: AudioSet\_SL, Freesound, SoundBible, and BBC Sound Effect. We use the entire AudioSet\_SL subset, exclude Freesound with inaccurate captions, and filter the SoundBible and BBC Sound Effect subsets by removing audios longer than 60 seconds. Ultimately, we select 121K audio clips from WavCaps. All audios are resampled to 16 kHz and we normalize all of them to 10 seconds by cropping or padding. In total, we compile 230K ten-second audio clips with class labels or captions as the source data for generating mixtures. 

For semantic separation, we employ triplet data (mixed audio, target audio, instruction) for supervised training. Instructions are generated using two task types and audio descriptions. We generate bidirectional instructions for each triplet and randomly select one during training, balancing the model’s ability to interpret both target and undesired audio events. \added{For example, for a mixture with two sources (A and B), we generate two types of instructions: (1) an extraction instruction targeting one source (e.g., “extract A”), and (2) a removal instruction targeting the other source (e.g., “remove B”).} Given that most source audios contain multiple audio events, when randomly selecting two samples to create each mixture, we follow the rule that the separated two audios do not have any overlap in audio classes. As text-audio pairs from various datasets feature diverse audio descriptions (e.g. FSD50K and ESC-50 provide different fixed sets of class labels, while AudioCaps, WavCaps and Clotho v2 offer detailed captions), it is necessary to unify the class labels to distinguish audio event types during audio mixing. We employ GPT-4\footnote{\url{https://chatgpt.com/}} \cite{achiam2023gpt} to semantically categorize different audio types of different datasets and summarize audio event tags from captions. For example, ``\textit{rain}" and ``\textit{raindrop}" are the same type and GPT-4 extracts tags ``\textit{man speech, thump, frog croak}" from the original caption ``\textit{An adult male speaks while thumps occur in the background, then frogs croak}." It is noted that these unified tags are only used in audio mixing to avoid mixing audios including the same categories. The instruction part will keep the original caption and labels to preserve the diversity. In total, we create 1.7M mixture-target audio pairs for training, containing approximately 4500 hours. 

\subsubsection{Q-Audio and AudioMAE}
To pretrain the Q-Audio module, we use the same datasets as that in Section \ref{sec:semantic extraction}, i.e. AudioCaps, ESC-50, Clotho v2, FSD50K and WavCaps, which provide detailed captions or class labels paired with each audio. Following the approaches of AudioMAE and MW-MAE \cite{yadav2023masked}, we train the MAE model using AudioSet-2M \cite{gemmeke2017audio}, encompassing both balanced and unbalanced subsets.

\subsubsection{Semantic-to-acoustic and acoustic decoder}
The decoder is trained using audios only without text labels. We gather data from AudioSet for general sounds, the Jamendo dataset for music \cite{bogdanov2019mtg}, and a multilingual speech dataset with 6k hours for clean speech \cite{jiang2024universal} that combines the Libri-Light medium subset \cite{kahn2020libri}, the multilingual speech subsets of the ICASSP 2021 DNS Challenge corpus \cite{reddy2021icassp}, and AVSpeech \cite{ephrat2018looking}. All audios are resampled to 16 kHz. We randomly sample 2,000 hours from each category. During TF-Codec training, audios are cropped to 3-second segments. For autoregressive audio token generation module, we set the maximum duration of each training audio to 10 seconds and perform padding masks on audio shorter than 10 seconds.

\subsection{Evaluation benchmark and metrics }
\subsubsection{Evaluation benchmark}
We compile evaluation data from the test sets of AudioCaps, FSD50K, and Clotho v2, for general audio assessment. The mixing strategy is the same as that used for training. As shown in Table \ref{tab:testsets}, we also create a ``3 Sets" that combine three datasets with an equal number of randomly sampled audio clips from each source dataset. Mixing sources from different datasets can account for distribution biases, leading to better approximations of real-world recordings. The average SNR between two components of the mixture is controlled within [-15dB, 15dB]. \added{It is noted that we utilize the simple ``extract ”and ``removal” instruction templates with original captions of the source datasets to ensure a fair comparison.}

\begin{table}[h]
    \centering
    \caption{Details of test sets}
    \label{tab:testsets}
    \begin{tabular}{lcccc}
        \hline
        Dataset & Num.sources & Caption & Duration(sec) & Num.pairs \\
        \hline         
        AudioCaps & 952 & \ding{51} & 10 & 500 \\
        Clotho & 1045 & \ding{51} & 15-30 & 500 \\
        FSD50K & 10231 & \ding{55} & 5-30 & 500  \\
        \hline
        3 Sets & 4000 & \ding{51} & 10 & 1000  \\
        \hline
    \end{tabular}
\end{table}

Additionally, we assess zero-shot audio separation performance on MUSIC dataset \cite{zhao2018sound}, which comprises 536 video recordings of individuals playing musical instruments across 11 classes (e.g., accordion, acoustic guitar, cello). 
Following \cite{dong2022clipsep}, we downloaded 46 video recordings from the test split, randomly selected two clips from different instrument classes and mixed them into 10-second segments, producing 500 audio pairs. The average SNR in mixing is within [-5dB, 5dB].  

\subsubsection{Objective metrics}
We evaluate our separation performance using log spectral distance (LSD), peak signal-to-noise ratio (PSNR), and Kullback-Leibler divergence (KL), following the AudioLDM\_eval audio generation evaluation pipeline\footnote{\url{https://github.com/haoheliu/audioldm_eval}}. LSD quantifies the difference between spectrograms of the predicted and target samples. PSNR measures the logarithmic ratio of the maximum possible signal power to the mean squared error between the predicted and target signals. KL divergence measures the similarity between the predicted and target audios with the probability distribution calculated by an audio tagging model. \added{While metrics like SI-SDR are common, they are highly sensitive to perceptually irrelevant artifacts, e.g. phase shifts or subtle but continuous hums, which can be introduced by generative decoders.}

To further evaluate the semantic correction of predicted audios, we leverage the CLAP score \cite{xiao2024reference}, which calculates audio-text similarity based on CLAP models \cite{elizalde2023clap,wu2023large}. The MSCLAP \cite{elizalde2023clap} is chosen for this measurement in our experiment. As CLAP tends to poorly capture temporal semantics and complex multi-source audio semantics \cite{wu2023audio}, we further introduce another semantic score, \textit{AFSim}, by calculating a cosine similarity between predicted and target signals on their semantic embeddings based on large audio language models (AudioLLM). Specifically, we leverage Audio Flamingo \cite{kong2024audio}, an audio LLM with advanced audio understanding capabilities, extract the feature from the penultimate (last second) layer by using the captioning prompt, and perform a mean-pooling to get the final embedding $AF \in \mathbb{R}^{1\times 2048}$, which is taken as the semantic embedding for our \textit{AFSim} score measurement. The superior semantic representation capability of the \deleted{$AF$}\added{$Audio\ Flamingo$} over CLAP is demonstrated in Table \ref{tab:audio_caption_qformer} in audio captioning evaluation.

\subsection{Implementation details}
For the Q-Audio module, we use a two-layer transformer with two self-attention layers and two cross-attention layers. The loss weights of the three losses are all set to 1.0. We choose FLAN-T5-base as the text encoder. The global-semantic separation module takes a 6-layer NAR transformer, with the Q-Audio audio and text encoders kept frozen during training. $\lambda_1$ and $\lambda_2$ are all set to 1.0.

We train the local-semantic separation module with a 12-layer NAR transformer which has 8 attention heads, and a hidden dimension of 768. During two-stage joint fine-tuning, we utilize the same training data and fintune the parameters of two NAR transformers, with other modules kept frozen. $P_{gt}$ is set to 0.1. The loss weights of $\mathcal{L}_{global}$ and $\mathcal{L}_{local}$ are set to 0.1 and 1.0, respectively.

The semantic-to-acoustic and acoustic decoder includes two parts. The autoregressive transformer has 12 layers with 8 attention heads and a hidden dimension of 1024. We take TF-Codec with a bitrate of 6 kbps in a causal setting, similar to that in \cite{jiang2024universal}.

\section{Results}
In this section, we evaluate the performance of our HSM-TSS approach and its modules including global-local representations and neural audio codec. \added{For all these comparisons of separation performance, we only use the simple ``extract" and ``removal" instruction templates as the compared methods are not designed for open vocabulary separation instruction inputs but only target audio captions.}

\subsection{Evaluation results on general audio}

\begin{table*}[t]
    \centering
    \caption{Objective and semantic evaluation of separation for general audio (3 Sets and Clotho). \addedII{``Ours\_local" denotes a single-separation-stage variant using only local-semantic separation with FLAN-T5 text encoder. ``Ours\_local*" denotes a single-separation-stage variant using only local-semantic separation with Q-Audio text encoder. No global-semantic separation is used in these two variants.}}
    \label{tab:extraction_semantic_3sets_clotho}
    \resizebox{\textwidth}{!}{ 
    \begin{tabular}{l c | ccc cc | ccc cc}
    \hline
    Model & Train.Data 
          & \multicolumn{5}{c|}{3 Sets} 
          & \multicolumn{5}{c}{Clotho} \\
    \cline{3-12}
          & (hrs) 
          & KL${(\downarrow)}$ & LSD${(\downarrow)}$ & PSNR${(\uparrow)}$ & AFSim${(\uparrow)}$ & CLAP${(\uparrow)}$ 
          & KL${(\downarrow)}$ & LSD${(\downarrow)}$ & PSNR${(\uparrow)}$ & AFSim${(\uparrow)}$ & CLAP${(\uparrow)}$ \\
    \hline
    GT &- & -& -& -& - &0.481 & -& -& -& - &0.437  \\
    \hline
    LASS-Net &17&2.577&3.170&16.42&0.615&0.268&2.713&1.696&18.21&0.590&0.314 \\
    CLIPSep&550&2.320&3.197&14.38&0.514&0.229&2.616&1.634&18.43&0.487&0.257 \\
    BiModalSep &670&1.634&3.118&19.49&0.678&0.420&1.819&1.613&19.71&0.674&0.385 \\    
    AudioSep &14100&1.027&3.037&22.50&0.730&0.357&1.191&1.616&21.57&0.720&\textbf{0.401} \\    
    \added{OmniSep} &\added{550}&\added{2.081}&\added{3.167}&\added{17.35}&\added{0.627}&\added{0.360}&\added{2.445}&\added{1.644}&\added{18.59}&\added{0.601}&\added{0.364} \\
    \added{FlowSep} &\added{1680}&\added{1.580}&\added{3.100}&\added{19.19}&\added{0.711}&\added{0.428}&\added{1.366}&\added{1.654}&\added{20.28}&\added{0.718}&\added{0.378} \\
    \hline
    \textbf{\textit{Extraction}}&&&&&&&&&&& \\
    Ours\_\added{local} &600&0.999&2.878&25.50&0.744&0.420&1.144&1.395&23.80&0.734&0.379 \\
    \added{Ours\_local*} &\added{600}&\added{0.997}&\added{2.879}&\added{25.51}&\added{0.741}&\added{0.421}&\added{1.135}&\added{1.383}&\added{23.95}&\added{0.730}&\added{0.381} \\
    Ours &600&\textbf{0.924}&\textbf{2.848}&\textbf{25.77}&\textbf{0.752}&\textbf{0.436}&\textbf{1.089}&1.378&\textbf{24.05}&\textbf{0.737}&0.393 \\
    \hline
    \textbf{\textit{Removal}}&&&&&&&&&&& \\
    Ours\_\added{local} &600&1.079&2.869&25.50&0.738&0.419&1.184&1.383&23.98&0.730&0.377 \\
    \added{Ours\_local*} &\added{600}&\added{1.061}&\added{2.866}&\added{25.53}&\added{0.734}&\added{0.420}&\added{1.182}&\added{1.381}&\added{23.99}&\added{0.728}&\added{0.376} \\
    Ours &600&1.007&2.852&25.66&0.741&0.428&1.112&\textbf{1.373}&23.94&0.733&0.380 \\
    \hline
    \end{tabular}
    }
\end{table*}

\begin{table*}[t]
    \centering
    \caption{Objective and semantic evaluation of separation for general audio (AudioCaps and FSD50K). \addedII{``Ours\_local" denotes a single-separation-stage variant using only local-semantic separation with FLAN-T5 text encoder. ``Ours\_local*" denotes a single-separation-stage variant using only local-semantic separation with Q-Audio text encoder. No global-semantic separation is used in these two variants.}}
    \label{tab:extraction_semantic_audiocaps_fsd50k}
    \resizebox{\textwidth}{!}{ 
    \begin{tabular}{l c | ccc cc | ccc cc}
    \hline
    Model & Train.Data 
          & \multicolumn{5}{c|}{AudioCaps} 
          & \multicolumn{5}{c}{FSD50K} \\
    \cline{3-12}
          & (hrs) 
          & KL${(\downarrow)}$ & LSD${(\downarrow)}$ & PSNR${(\uparrow)}$ & AFSim${(\uparrow)}$ & CLAP${(\uparrow)}$ 
          & KL${(\downarrow)}$ & LSD${(\downarrow)}$ & PSNR${(\uparrow)}$ & AFSim${(\uparrow)}$ & CLAP${(\uparrow)}$ \\
    \hline
    GT &- & -& -& -& - &0.575 & -& -& - & - &0.466\\
    \hline
    LASS-Net &17&2.446&2.276&17.64&0.641&0.264&2.474&4.528&15.17&0.592&0.223 \\
    CLIPSep&550&2.738&2.464&17.07&0.519&0.214&2.967&4.475&17.15&0.525&0.203 \\
    BiModalSep &670&1.441&2.256&19.44&0.713&0.501&1.789&4.428&18.89&0.631&0.387 \\    
    AudioSep &14100&1.002&2.142&21.17&0.752&0.376&1.172&4.313&23.66&0.699&0.305 \\    
    \added{OmniSep} &\added{550}&\added{2.174}&\added{2.401}&\added{18.03}&\added{0.640}&\added{0.403}&\added{1.897}&\added{4.398}&\added{19.25}&\added{0.627}&\added{0.326} \\
    \added{FlowSep} &\added{1680}&\added{1.238}&\added{2.012}&\added{20.37}&\added{0.751}&\added{0.508}&\added{1.683}&\added{4.335}&\added{19.27}&\added{0.682}&\added{0.425} \\
    \hline
    \textbf{\textit{Extraction}}&&&&&&&&&&& \\
    Ours\_\added{local} &600&0.960&1.916&23.49&0.763&0.500&0.948&4.197&27.11&0.721&0.428 \\
    \added{Ours\_local*} &\added{600}&\added{1.034}&\added{1.895}&\added{23.71}&\added{0.742}&\added{0.498}&\added{1.068}&\added{4.209}&\added{26.65}&\added{0.720}&\added{0.430} \\
    Ours &600&\textbf{0.910}&\textbf{1.884}&\textbf{23.93}&\textbf{0.767}&\textbf{0.511}&\textbf{0.889}&\textbf{4.156}&\textbf{27.34}&\textbf{0.729}&\textbf{0.437} \\
    \hline
    \textbf{\textit{Removal}}&&&&&&&&&&& \\
    Ours\_\added{local} &600&1.148&1.919&23.66&0.755&0.486&1.069&4.209&26.65&0.718&0.420 \\
    \added{Ours\_local*} &\added{600}&\added{1.152}&\added{1.917}&\added{23.71}&\added{0.754}&\added{0.486}&\added{1.061}&\added{4.199}&\added{26.74}&\added{0.715}&\added{0.421} \\
    Ours &600&1.105&1.908&23.88&0.763&0.492&1.041&4.179&27.00&0.720&0.425 \\
    \hline
    \end{tabular}
    }
\end{table*}

        

We compare the separation performance of our HSM-TSS with several text-queried audio separation methods, LASSNet \cite{liu2022separate}, AudioSep \cite{liu2024separate}, CLIPSep \cite{dong2022clipsep}, OmniSep \cite{cheng2024omnisep} and BiModalSep \cite{mahmud2024weakly}, which are all frequency-domain mask-based approaches. We also compare with a generative model, FlowSep \cite{yuan2025flowsep}. LASS-Net employs a pre-trained BERT as the text query encoder and ResUNet as the separation model, while AudioSep further integrates a CLAP text encoder as the query network and trains on a much larger dataset, yielding substantial performance gains. CLIPSep adopts CLIP as the query encoder and a Sound-of-Pixels (SOP)-based separation model, trained on approximately 500 hours of noisy audio-visual data from the VGGSound \cite{chen2020vggsound} dataset using hybrid vision-text supervision. BiModalSep introduces a weakly-supervised approach and leverages bi-modal semantic similarity via CLAP to align single-source language prompts with audio predictions, achieving robust separation without curated single-source data. \added{OmniSep leverages multi-modal queries flexibly through a omni-modal encoder ImageBind \cite{girdhar2023imagebind}. FlowSep utilizes rectified flow matching for separation. } We evaluate their performance using their official open-source models on our benchmarks. 

To show the effectiveness of our hierarchical modeling, we also compare with a single-separation-stage variant of our HSM-TSS, termed \deleted{``Ours\_single"}\added{``Ours\_local"} in Table \ref{tab:extraction_semantic_3sets_clotho} \added{and Table \ref{tab:extraction_semantic_audiocaps_fsd50k}}. It removes the first global-semantic separation stage and only leverages the local-semantic separation stage. The text encoder is FLAN-T5 without Q-Audio in this variant. \added{Additionally, we present an ablation study by evaluating another single-separation-stage variant which leverages Q-Audio text encoder, noted as ``Ours\_local*".} We provide results for not only ``extraction" as used in \deleted{four}\added{six} compared methods but also ``removal" task types that produce the same target audios. 

As shown in Table \ref{tab:extraction_semantic_3sets_clotho} \added{and Table \ref{tab:extraction_semantic_audiocaps_fsd50k}}, our variant ``Ours\_local" outperforms baselines in several metrics by a large margin. With the introduction of another global-semantic separation, the proposed HSM-TSS, noted as ``Ours", outperforms both baseline methods and ``Ours\_local" across various metrics. When compared to \deleted{four}\added{other} methods, LASSNet, CLIPSep, AudioSep, \added{OmniSep, FlowSep }and BiModalSep, our approach consistently achieves higher scores in the three \added{objective} metrics, especially for the 3 Sets benchmark, highlighting its effectiveness in separating target audio events from complex mixtures. In contrast to the single-separation-stage setting, our dual-stage separation allows for progressive refinement of audio features from coarse to fine, leading to more precise separation of target sounds\deleted{, as demonstrated in Table \ref{tab:extraction_semantic_3sets_clotho}}\added{, e.g. KL divergence decreases from 0.999 to 0.924 on 3 Sets}. We can also see that our model with ``removal" instruction achieves comparable scores with ``extraction", showing the bidirectional operation capability of our method. 

Table \added{\ref{tab:extraction_semantic_3sets_clotho} and Table \ref{tab:extraction_semantic_audiocaps_fsd50k}} \added{also show the evaluation of semantic correctness by CLAP score and AFSim.} \deleted{show the evaluation results on semantic consistency of the output audios with instructions.} For ``removal" setting, we use the same audio description as ``extraction" to calculate the CLAP score. We can see that our method consistently outperforms other methods and the single-separation-stage variants in both AFSim and CLAP scores, showing its superior capability to follow instructions with diverse audio descriptions. \deleted{For all these comparisons, we only use the simple ``extract" and ``removal" instruction templates as the compared methods are not designed for open vocabulary separation instruction inputs but only target audio captions.} Our method has good arbitrary instruction following capability, as shown in our demo page\footnote{\url{https://hsm-tss.github.io}}.

\added{It is also observed that the impact of using the Q-Audio text encoder in the single-separation-stage model is marginal. The performance of ``Ours\_local*" and \deleted{``Ours\_single"}\added{``Ours\_local"} remains nearly identical across multiple evaluation metrics, and both are consistently inferior to ``Ours". This is reasonable as Q-Audio is designed to align text and audio in the shared space, but in local-semantic-separation stage, the audio features are from AudioMAE instead of Q-Audio. Further extracting text feature from FLAN-T5 does not alter its semantic representativeness.}



\subsection{Zero-shot performance on unseen datasets}
We perform zero-shot evaluation on mixtures of music instruments from the MUSIC dataset, as we do not use any specialized music data during training. We can see from Table \ref{tab:music} that our method outperforms all other methods in all metrics. It's worth noting that the audio clips with clear musical instrument labels in our training data primarily come from FSD50K, accounting for less than 5\% of the entire dataset, significantly lower than the proportion in AudioSep's training dataset, AudioSet and VGGSound, that contain rich YouTube-sourced music instrument data. The superior performance of our HSM-TSS method demonstrates its strong generalization capability across diverse music content.

\begin{table}[t]
    \centering
    \caption{Evaluation on zero-shot MUSIC dataset. \addedII{``Ours\_local" denotes the single-separation-stage variant using only local-semantic separation.}}
    \label{tab:music}
    \resizebox{\columnwidth}{!}{ 
    \begin{tabular}{lcccccc}
        \hline
        Model & KL${(\downarrow)}$ & LSD${(\downarrow)}$ & PSNR${(\uparrow)}$ &AFSim${(\uparrow)}$ &CLAP${(\uparrow)}$ \\
        \hline
        GT &- &- &- &- &0.490   \\
        \hline
        LASS-Net & 4.180 & 2.066 & 13.23 &0.503 &0.126  \\
        CLIPSep & 2.179 & 2.554 & 14.07 &0.657&0.372 \\
        BiModalSep & 2.607 & 2.250 & 14.35 &0.671&0.374 \\
        AudioSep & 0.535 & 1.624 & 22.01 &0.804&0.342 \\        
        \added{OmniSep} & \added{1.200} & \added{2.161} & \added{16.91} &\added{0.716}&\added{0.383} \\
        \added{FlowSep} & \added{1.659} & \added{1.706} & \added{18.20} &\added{0.733}&\added{0.437} \\
        \hline
        \textbf{\textit{Extraction}}&&&&& \\
        Ours\_\added{local} &0.585&1.400&23.22&0.799&0.463 \\
        Ours &\textbf{0.501} &\textbf{1.375} &\textbf{23.85} &\textbf{0.811}&\textbf{0.470} \\
        \hline
    \end{tabular}
    }
\end{table}

\begin{addedblockII}
\subsection{Ablation study on local-semantic representation}
To further investigate the contribution of AudioMAE as the local-semantic representation, we conduct additional ablation experiments under the single-separation-stage setting without a global-semantic separation. As shown in Table \ref{tab:audiomae_ablation}, we compare ``Ours\_local" with two other variants, ``Ours\_local\_CLAP" and ``Ours\_acoustic". In ``Ours\_local\_CLAP", we replace AudioMAE with the last-layer local features from the CLAP audio encoder \cite{wu2023large} before pooling. Although CLAP features exhibit strong semantic alignment with text, they tend to compromise local acoustic details for reconstruction. The ``Ours\_acoustic" omits AudioMAE as the local semantic separation and directly utilizes neural codec embeddings for separation by a transformer. While these embeddings retain detailed acoustic information, they lack semantic abstraction, which makes it difficult to separate target signals following text instructions. As shown in Table \ref{tab:audiomae_ablation}, both alternatives yield much inferior results compared to ``Ours\_local" using AudioMAE across all metrics. These findings confirm that neither approach can capture fine-grained acoustic details and semantic relevance as effectively as AudioMAE.

\begin{table}[t]
    \centering
    \caption{\addedII{Ablation results of AudioMAE on 3 Sets. ``Ours\_local" denotes the single-separation-stage variant using AudioMAE for local-semantic separation. ``Ours\_local\_CLAP" denotes the single-separation-stage variant using CLAP encoder feature for local-semantic separation. ``Ours\_acoustic" denotes the single-acoustic-separation variant performing separation on acoustic features only without AudioMAE. }}
    \label{tab:audiomae_ablation}
    \resizebox{\columnwidth}{!}{ 
    \begin{tabular}{lcccccc}
        \hline
        Model & KL${(\downarrow)}$ & LSD${(\downarrow)}$ & PSNR${(\uparrow)}$ &AFSim${(\uparrow)}$ &CLAP${(\uparrow)}$  \\
        \hline
        GT &- &- &- &- &0.481   \\
        \hline
        \textbf{\textit{Extraction}}&&&&& \\        
        Ours\_acoustic (w/o AudioMAE) & 2.024 & 3.106 & 16.59 & 0.601 & 0.293  \\
        Ours\_local\_CLAP & 1.901 & 3.002 & 17.95 & 0.634 & 0.310  \\          
        Ours\_local & 0.999 & 2.878 & 25.50 & 0.744 & 0.420    \\
        \hline
    \end{tabular}
    }
\end{table}
\end{addedblockII}

\begin{addedblock}
\subsection{Impact of source overlap}
To assess the robustness of the proposed model under varying levels of difficulty, we perform an ablation study by adjusting the \textit{overlap ratio} between two source signals. Figure \ref{fig:overlap} illustrates the separation results in terms of KL divergence and AFSim across different overlap levels. 

It can be seen that as the overlap ratio increases, the separation task becomes more challenging. However, our model consistently outperforms the baselines. More importantly, the performance margin becomes more pronounced under higher overlap ratios, where our method achieves the largest relative improvement over both ``Ours\_local" and ``AudioSep". These findings demonstrate that the proposed framework is particularly effective in handling complex mixtures with heavily overlapped sources, highlighting its robustness in real-world acoustic scenarios.

\begin{figure}[t]
    \centering
    \includegraphics[width=1.0\columnwidth]{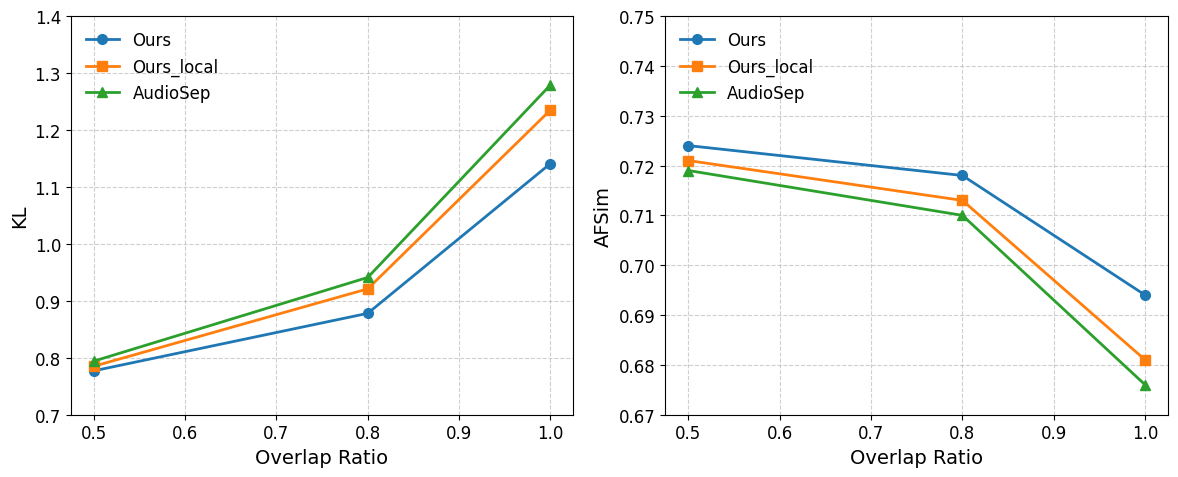}
    \caption{The trends of KL and AFSim with varying source overlap ratios.}
    \label{fig:overlap}
\end{figure}
\end{addedblock}

\subsection{Visualization}
\subsubsection{t-SNE visualization}
\begin{figure}[t]
    \centering
    \includegraphics[width=0.9\columnwidth]{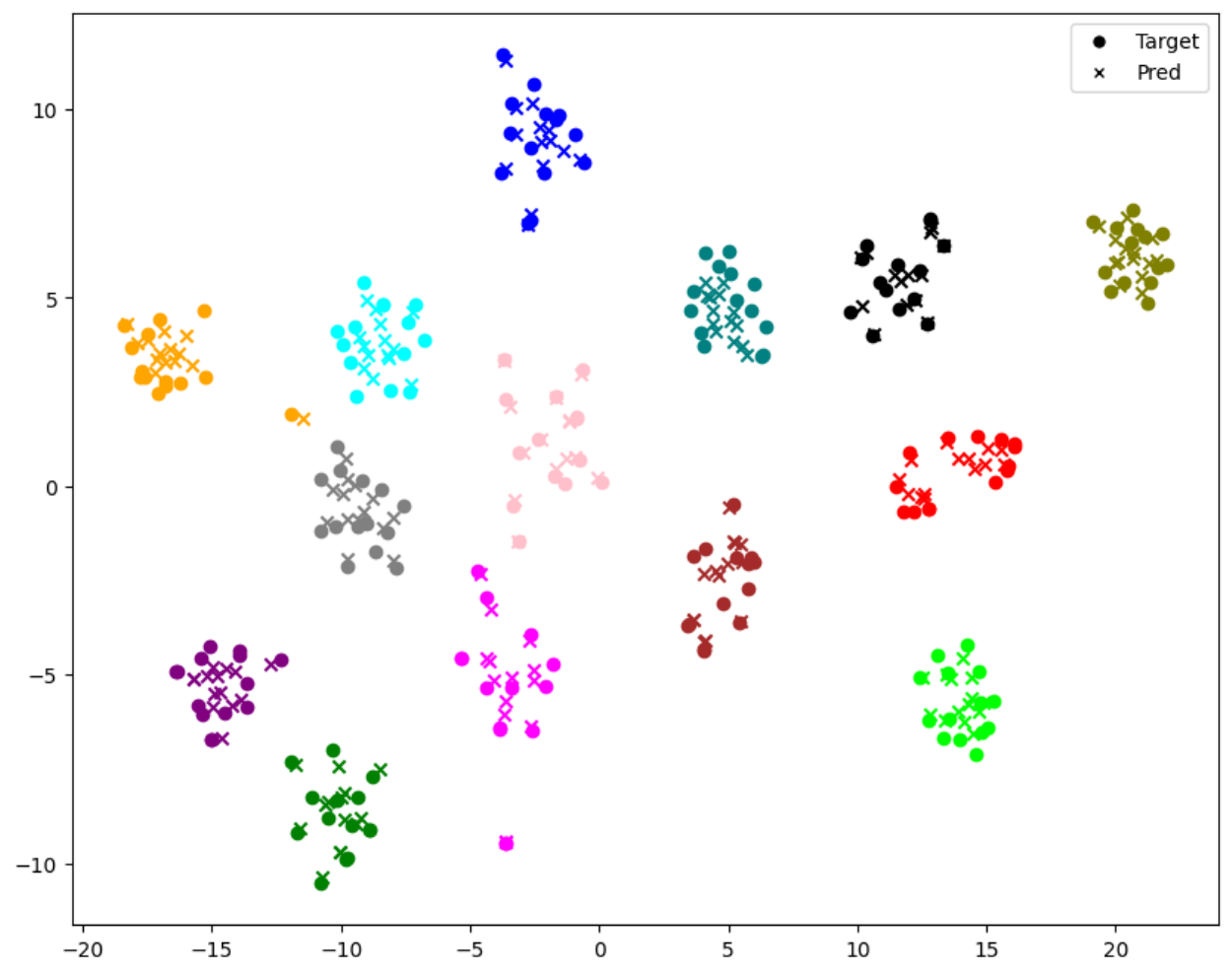}
    \caption{t-SNE visualization of global-semantic features: ground-truth vs. predicted.}
    \label{fig:tsne}
\end{figure}

To show how the global-semantic separation performs, we visualize the extracted features from this stage with t-SNE \cite{van2008visualizing}. In Figure \ref{fig:tsne}, each color shows a sound event class and we present the ground-truth global audio semantic feature and the separated output with different markers. It can be observed that our global-semantic separation model effectively extracts target audio features with good clustering.

\subsubsection{Visualization of attention maps}
\begin{figure}[t]
    \centering
    \includegraphics[width=\columnwidth]{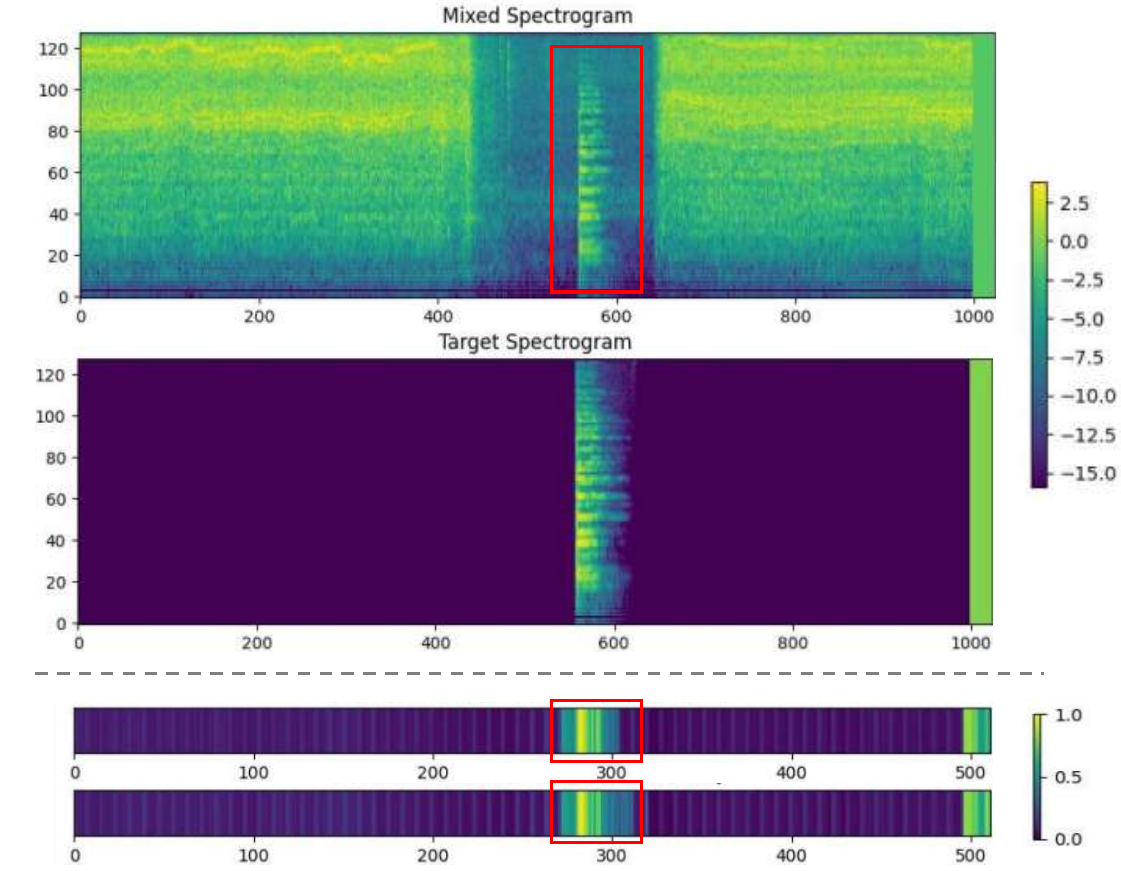}
    \caption{Visualization of attention weights. The upper subfigures are the spectrograms of the mixed and target audios, respectively. The bottom two subfigures show the attention weights of the global audio feature attending to local AudioMAE patches for ground-truth and predicted global feature inputs, respectively. }
    \label{fig:attention_map}
\end{figure}

To show how the predicted global semantic feature helps the local-semantic separation stage, we visualize the attention map from the last layer of the NAR transformer. In Figure \ref{fig:attention_map}, the bottom two subfigures show the attention weights of the global feature attending to all 512 local semantic patches in temporal order, with the upper one takes ground-truth global feature as input and the bottom one uses previous global-semantic separation stage output feature. We can observe that the high peaks happen where target audio event occurs, which well demonstrates the semantic alignment of predicted global feature with target audio. 

\subsubsection{Visualization of separation outputs}
\begin{figure*}[t]
    \centering
    \includegraphics[width=\textwidth]{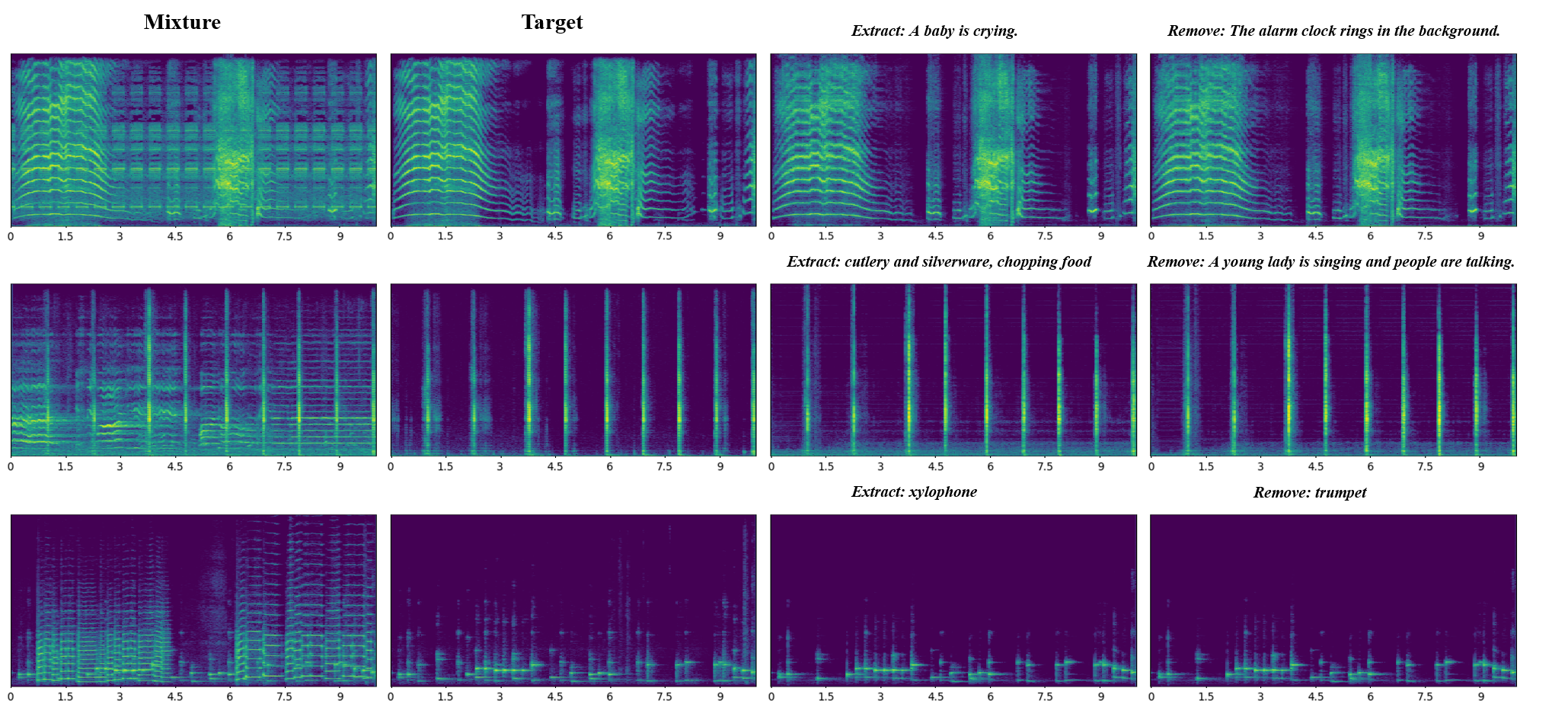}
    \caption{Visualization of mixture, target and the separated audio by our method. }
    \label{fig:visualize_spectra}
\end{figure*}

We show the spectrograms for mixture input, target audio and the separated audio with text queries for both ``extraction" and ``removal" commands in Figure \ref{fig:visualize_spectra}. We can see that the spectrograms from separated output audios closely resemble the target one, and the two commands achieve similar results, consistent with our objective evaluation results. 

\subsection{Semantic representation evaluation}
\subsubsection{Global semantic feature evaluation}\label{sec:downstream}
To evaluate the cross-modal alignment and semantic representation capbility of our Q-Audio, we follow the settings of LAION-CLAP \cite{wu2023large} on AudioCaps and Clotho datasets. Table \ref{tab:text2audio_retrieval_combined} shows the text-to-audio retrieval results. Compared to the widely used LAION-CLAP \cite{wu2023large} and MSCLAP \cite{elizalde2023clap}, Q-Audio achieves better performance, underscoring its capability in audio-language modeling. Table \ref{tab:audio_caption_qformer} shows the audio captioning evaluation, where a linear mapping layer and a pretrained GPT2 \cite{radford2019language} are used as the downstream model to generate audio captions. Here we only train the downstream model and keep audio feature extraction frozen. Although our Q-Audio is not pretrained on the captioning task as CLAP models, it still achieves higher scores on the three metrics, showing its excellent semantic representation capability.  

\begin{table}[h]
    \centering
    \caption{Text-to-audio retrieval performance}
    \label{tab:text2audio_retrieval_combined}
    \begin{tabular}{lccc|ccc}
        \hline
        Model & \multicolumn{3}{c|}{Clotho} & \multicolumn{3}{c}{AudioCaps} \\
        \cline{2-4} \cline{5-7}
        & R@1 & R@5 & R@10 & R@1 & R@5 & R@10 \\
        \hline
        LAION-CLAP & 14.6 & 37.3 & 49.9 & 20.4 & 49.7 & 64.3\\
        MSCLAP & 15.6 & 38.6 & 51.3 & 25.4 & 55.6 & 69.7 \\
        Q-Audio & \textbf{17.1} & \textbf{41.4} & \textbf{54.1} & \textbf{26.6} & \textbf{56.8} & \textbf{70.4}  \\
        \hline
    \end{tabular}
\end{table}

\begin{table}[h]
    \centering
    \caption{Audio captioning on AudioCaps and Clotho}
    \label{tab:audio_caption_qformer}
    \begin{tabular}{lcccc}
        \hline
        Model & METEOR${(\uparrow)}$ & SPICE${(\uparrow)}$ & SPIDEr${(\uparrow)}$ \\
        \hline
        \added{Audio Flamingo} & \textbf{0.121} & \textbf{0.170} & \textbf{0.336} \\
        LAION-CLAP & 0.115 & 0.152 & 0.311 \\
        MSCLAP & 0.116 & 0.160 & 0.326 \\
        Q-Audio & \textbf{0.118} & \textbf{0.164} & \textbf{0.329} \\
        \hline
    \end{tabular}
\end{table}

\subsubsection{Local semantic feature evaluation}
We evaluate our pretrained AudioMAE model on the HEAR benchmark \cite{turian2022hear} and three tasks, ESC-50, SC-5h and NS-5h, where only the downstream model is trained. As shown in Table \ref{tab:hear_benchmark}, our pretrained model achieves good performance, outperforming the official AudioMAE-B \cite{huang2022masked}. \added{This improvement can be attributed to two designs inspired by MW-MAE \cite{yadav2023masked}, i.e. employing a smaller decoder to encourage the encoder to capture richer contextual representations, and removing the dataset-level normalization.}

\begin{table}[h]
    \centering
    \caption{Evaluation of AudioMAE on HEAR benchmark}
    \label{tab:hear_benchmark}
    \hspace{-0.4cm}
    \begin{tabular}{lccc}
        \hline
        Model & ESC-50${(\uparrow)}$ & SC-5h${(\uparrow)}$ & NS-5h${(\uparrow)}$ \\
        \hline
        AudioMAE-B \cite{huang2022masked} & 57.6 & 33.9 & 61.4  \\
        Our pretrained AudioMAE & \textbf{83.7} & \textbf{77.1} & \textbf{66.0} \\
        \hline        
    \end{tabular}    
\end{table}

\subsection{Neural codec performance}
In Table \ref{tab:codec_metrics}, we assess the reconstruction quality of general audio by our neural codec TF-Codec. We randomly sample 1000 audios from AudioSet validation set for evaluation. All models have a bitrate of 6 kbps. We can see that although our general audio TF-codec has a low latency by using a causal structure and much less parameters, it achieves good performance, superior than EnCodec \cite{defossez2022high}. DAC \cite{kumar2023high} achieves the best performance but it is much heavier with a non-causal structure. It should be noted that we utilize TF-Codec just as a proof of concept, and future works may utilize any non-causal codec with superior performances. 

\begin{table}[h]
    \centering
    \caption{Neural codec performance on general audio at 6kbps}
    \scriptsize
    \label{tab:codec_metrics}
    \begin{tabular}{lccccc}
        \hline
        Model &\#param & Causal & MEL-D${(\downarrow)}$ & LSD${(\downarrow)}$ & VISQOL${(\uparrow)}$ \\
        \hline
        EnCodec &14.85M & \ding{51} & 1.047 & 4.058 & 4.209 \\
        DAC &74.17M & \ding{55} & 0.630 & 3.556 & 4.521\\
        \hline
        Our TF-Codec & 7.63M &\ding{51} & 0.797 & 3.588& 4.375 \\
        \hline
    \end{tabular}
\end{table}


\begin{addedblock}
\subsection{Complexity}
We compare the computational complexity with UNet-style separation method AudioSep and generative model based approach FlowSep in Table \ref{tab:complexity}. Here “\# Param\_sep" and “\# Param\_pretrained" denote the model parameters for separation module and other pretrained text or audio encoder/decoders, respectively. It is noted that although our HSM-TSS has more parameters than AudioSep, its complexity in number of FLOPs is much lower than both AudioSep and FlowSep as AudioSep uses a heavy ResUNet module for separation and FlowSep utilizes several steps, default to 20 in Table \ref{tab:complexity}, to generate the separated audio feature. We utilize \textit{fvcore\footnote{\url{https://github.com/facebookresearch/fvcore}}} to calculate the FLOPs for each model.
\end{addedblock}

\begin{table}[h]
    \centering
    \caption{\added{Complexity comparisons for a 10-second audio}}
    \label{tab:complexity}
    \begin{tabular}{lccc}
        \hline
        \added{
        Model} & \added{FLOPs/G} & \added{\# Param\_sep/M} & \added{\# Param\_pretrained/M} \\
        \hline
        \added{AudioSep} &\added{333.7} & \added{39.0} & \added{199.6} \\
        \added{FlowSep} & \added{2046.9} & \added{265.6} & \added{410.5} \\
        \hline
        \added{HSM-TSS} &\added{183.8} &\added{129.9} & \added{378.3}\\
        \hline
    \end{tabular}
\end{table}

\section{Conclusion and future works}
In this study, we propose a hierarchical modeling and separation framework for text-queried audio source separation, decoupling multi-level semantic feature separation and acoustic reconstruction. Leveraging Q-Audio for global-semantic modeling on top of AudioMAE for structure-preserving representations, our model achieves superior separation performance and semantic correctness, outperforming existing methods. \deleted{The instruction parser enhances flexibility in handling diverse text queries with frozen LLMs.} In future work, we will scale up the model in training data coverage,  which incorporates speech as well, and explore \deleted{more fantastic editing options besides separation.}\added{a wider range of instruction-based editing capabilities beyond separation, such as style transfer or in-place modification.}

\bibliographystyle{IEEEtran}
\bibliography{TASLP/reference}



\end{document}